\documentclass[aip,pop,numerical,nofootinbib,reprint,showpacs,showkeys]{revtex4-1}

\usepackage{amsmath}
\usepackage{amsfonts}
\usepackage{amssymb}
\usepackage{graphicx}
\usepackage{leftidx}

\begin{document}

\newcommand{\bo}{\ensuremath{\boldsymbol{B}_0}}
\newcommand{\dm}{\ensuremath{D_{\mu\mu}}}
\newcommand{\const}{\ensuremath{\text{const}}}
\newcommand{\sgn}{\ensuremath{\text{sgn}\,}}
\newcommand{\abs}[1]{\ensuremath{\left|#1\right|}}
\newcommand{\erf}[2][]{\ensuremath{\text{erf#1}\!\left(#2\right)}}
\newcommand{\kB}{\ensuremath{k_{\text B}}}
\newcommand{\eps}{\ensuremath{\varepsilon}}
\newcommand{\vA}{\ensuremath{v_{\mathrm A}}}

\newcommand{\lD}{\ensuremath{\lambda_{\text D}}}

\newcommand{\be}{\begin{equation}}
\newcommand{\ee}{\end{equation}}
\newcommand{\bs}{\begin{subequations}}
\newcommand{\es}{\end{subequations}}
\def \bdm{\begin{eqnarray}}
\def \edm{\end{eqnarray}}

\newcommand{\ex}{\ensuremath{\hat{\boldsymbol{e}}_x}}
\newcommand{\ey}{\ensuremath{\hat{\boldsymbol{e}}_y}}
\newcommand{\ez}{\ensuremath{\hat{\boldsymbol{e}}_z}}

\newcommand{\R}{\ensuremath{\mathds{R}}}

\newcommand{\pa}{\ensuremath{_\parallel}}
\newcommand{\se}{\ensuremath{_\perp}}

\newcommand{\De}{\ensuremath{\varDelta}}
\newcommand{\Om}{\ensuremath{\varOmega}}
\newcommand{\Ph}{\ensuremath{\varPhi}}
\newcommand{\Ga}{\ensuremath{\varGamma}}
\newcommand{\La}{\ensuremath{\varLambda}}
\newcommand{\Ps}{\ensuremath{\varPsi}}
\newcommand{\Up}{\ensuremath{\varUpsilon}}

\newcommand{\uint}{\ensuremath{\int_{-\infty}^\infty}}
\renewcommand{\iint}{\ensuremath{\int\!\!\!\!\int}}
\renewcommand{\iiiint}{\int\!\!\!\!\int\!\!\!\!\int\!\!\!\!\int}
\newcommand{\usum}{\ensuremath{\sum_{n=-\infty}^\infty}}
\newcommand{\esum}{\ensuremath{\sum_{n=1}^\infty}}
\newcommand{\nsum}{\ensuremath{\sum_{n=0}^\infty}}
\newcommand{\Ff}[1]{\ensuremath{\leftidx{_2}{F}{_2}\!\left({\displaystyle #1}\right)}}

\newcommand{\f}[1]{\ensuremath{\boldsymbol{#1}}}
\newcommand{\m}[1]{\ensuremath{\left\langle #1\right\rangle}}
\newcommand{\im}[1]{\ensuremath{\mathfrak{I\!m}\left(#1\right)}}

\newcommand{\pd}[2][]{\ensuremath{\frac{\partial #1}{\partial #2}}}
\newcommand{\dd}[2][]{\ensuremath{\frac{\mathrm{d} #1}{\mathrm{d} #2}}}
\newcommand{\df}{\ensuremath{\mathrm{d}}}

\newcommand{\etal}{ \emph{et~al.}}
\renewcommand{\thefootnote}{\fnsymbol{footnote})}
\setcounter{footnote}{2}

\title{On numerical turbulence generation for test-particle simulations}

\author{R.\,C. Tautz}
\email{rct@gmx.eu}
\affiliation{\mbox{Zentrum f\"ur Astronomie und Astrophysik, Technische Universit\"at Berlin,}\\Hardenbergstra\ss{}e 36, D-10623 Berlin, Germany}
\author{A. Dosch}
\email{alexanderm.dosch@gmail.com}
\affiliation{\mbox{Center for Space Plasmas and Aeronomic Research, University of Alabama in Huntsville},\\320~Sparkman Drive, Huntsville, AL~35805, USA}

\date{\today}

\begin{abstract}
A modified method is presented to generate artificial magnetic turbulence that is used for test-particle simulations. Such turbulent fields are obtained from the superposition of a set of wave modes with random polarizations and random directions of propagation.  First, it is shown that the new method simultaneously fulfils requirements of isotropy, equal mean amplitude and variance for all field components, and vanishing divergence. Second, the number of wave modes required for a stochastic particle behavior is investigated by using a Lyapunov approach. For the special case of slab turbulence, it is shown that already for 16 wave modes the particle behavior agrees with that shown for considerably larger numbers of wave modes.
\end{abstract}

\pacs{52.25.Dg --- 52.25.Xz --- 52.65.Ff --- 96.50.S- --- 96.50.Vg}
\keywords{plasmas --- turbulence --- diffusion --- transport --- cosmic rays}

\maketitle

\section{Introduction}\label{intro}

Although it appears to be a simple problem, the question of how energetic charged particles are scattered in turbulent electromagnetic fields is still far from being solved. Over the years, a variety of analytical theories\cite{rs:rays,mat03:nlg,sha09:nli,sha10:uni} and (semi-)numerical methods\cite{ach92:acc,gia99:sim,ruf12:nlg} has been used. Considerable progress has been made toward understanding specific scenarios such as cosmic-ray diffusion in the Solar system.\cite{par65:pas,tau12:pal} While numerical simulations can relatively easy be extended for other parameter regimes, such is considerably more difficult for analytical calculations,\cite{sha05:soq,tau08:soq} which have to take into account different energy regimes, different turbulence models, and different time dependences of the turbulence. Furthermore, wave-particle interactions also play a role in the generation of strong magnetic turbulence via plasma instabilities and subsequent particle radiation,\cite{sch05:ori,tau12:rad} both in the astrophysical context as well as in laboratory plasmas and fusion devices. However, here the focus will be set on test particles, which have a sufficiently low density so that the fields remain unaffected. Self-consistent calculations have been done as well, which however focus on qualitative results based on non-linear perturbation approaches\cite{car91:mot} or on the combination of mostly irreconcilable ansatzes.\cite{tau08:wei}

As a basic test case, much effort has been focused on magnetostatic fields,\cite{jok66:qlt,tau06:sta} which conserve the kinetic energy and where only the particles' direction of motion is changed through scattering processes. As a numerical counterpart to analytical calculations, special attention has been diverted to a Monte-Carlo method, which traces the trajectories of test particles in a prescribed artificial turbulence. In the case of the \text{Padian} code,\cite{tau10:pad} normalized variables are introduced as $\tau=\Om t$, where $\Om=qB_0/(mc)$ is the gyrofrequency, and $\f R=\gamma\f v/(\Om\ell_0)$, which is named a normalized rigidity with $\ell_0$ the turbulence bend-over scale. The equations of motion then read
\bs
\begin{align}
\partial\left(\f x/\ell_0\right)/\partial\tau&=\f R\\
\partial\f R/\partial\tau&=\f R\times\left[\f{\hat e}_{B_0}+\left(\delta B/B_0\right)\f{\hat e}_{\delta B}\right],
\end{align}
\es
where $\f{\hat e}_{B_0}$ and $\f{\hat e}_{\delta B}$ are unit vectors denoting the directions of the mean and turbulent magnetic fields, respectively. While the background field is usually homogeneous (sometimes, special geometries are assumed such as the Parker spiral\cite{par58:spi,tau11:spi} or adiabatically focused fields\cite{roe69:int,tau12:adf}), the spatial (and temporal, for example due to plasma waves\cite{tau10:wav}) variation of the turbulent fields are considerably more complicated.

Two basic onsets have been used over the years: First, the \emph{grid approach} discretizes both the spatial and the wavenumber coordinates to enable a fast Fourier transform. Before integrating particle trajectories, the turbulent fields are specified on the grid using a superposition of (standing) wave modes. The advantage is that the turbulence structure has to be evaluated only once, while, on the negative side, up to $2^{22}$ Fourier modes\cite{qin02b:sim,ruf06:flr,ruf08:per} have to be used in order to obtain a sufficient resolution of spatial structures.

Second, following \citeauthor{gia99:sim},\cite{gia99:sim} (which will hereafter be denoted as GJ99) the \emph{continuous approach} is based on a just-in-time evaluation of the turbulent fields at each particle position, which has the advantage that no grid structure is needed. For the test case of isotropic turbulence, the three basic requirements for the turbulence generator are: (i) all three turbulent magnetic field components have the same magnitude on average; (ii) the probability for the random orientation of the sample Fourier modes is isotropic; and (iii) the divergence of the turbulent magnetic field must be zero. As shown recently,\cite{tau12:sim} this was not possible using previously used approaches because at least one condition was always violated. Therefore, we feel that it is appropriate to present a modification of the original method that fulfills all three conditions.

A third, relatively novel ansatz is based on the application of stochastic differential equation solvers\cite{gar09:sto,kop12:sto} for the cosmic-ray transport equation. The method has been applied to pick-up ions, particle acceleration, and cosmic-ray diffusion.\cite{ach92:acc,cha95:pic,fic96:pic,eff11:pro,str11:ene,eff12:ani}

For the two first approaches, there appear to be no systematic investigations of the number of wave modes that is necessary in order to have a sufficiently turbulent electromagnetic field structure. While, from a basic point of view, isotropic and homogeneous turbulence is obtained only in the limit of infinitely many wave modes,\cite{bat82:tur} there are indications that anisotropic structures might be obliterated already for a finite number of wave modes.\footnote{F. Alouani-Bibi, private communication (2012).} In the literature, various numbers of wave modes have been used with seemingly no systematic motivation besides a fondness for even numbers (see Table~\ref{ta:lit}). In the \textsc{Padian} code, usually 128--2048 modes are used.

\begin{table}[t]
\begin{center}
\begin{tabular}{lll}
\multicolumn{3}{l}{Table I. An incomplete survey of wave modes and}\\
\multicolumn{3}{l}{simulations used type in the literature}\\[7pt]\hline\hline
Reference No. 							& Simulation type	& Number of wave modes\\[4pt]\hline
\citenum{ost93:mom}						& Continuous		& 18\\
\citenum{mic96:alf}						& Continuous		& 24\\
\citenum{gia94:mul}						& Continuous		& 50\\
\citenum{ott92:sca}						& Continuous		& 64\\
\citenum{can04:hig}						& Continuous		& 100\\
\citenum{lai12:str}						& Continuous		& 128--1024\\
\citenum{mic98:obl}						& Continuous		& 384\\
\citenum{mic01:sim}						& Continuous		& 768\\
\citenum{rev08:tra}						& Continuous		& 5000\\
\citenum{reu96:per}						& Grid			& 7168\\
\citenum{zim09:per}						& Grid			& 8200\\
\citenum{zim05:ano,zim06:dif}				& Grid			& $\sim10^6$\\
\citenum{mac00:wea}						& Grid			& 1,657,864\\
\citenum{qin02b:sim,ruf06:flr,ruf08:per}	& Grid			& $2^{21}$--$2^{22}$\\[4pt]\hline\hline
\end{tabular}
\end{center}
\label{ta:lit}
\end{table}

The purpose of this article is thefore two-fold: In Sec.~\ref{rot}, a new method for the generation of turbulent electromagnetic fields is presented and the properties of the resulting diffusion coefficients are discussed for the special case of isotropic turbulence both by direct evaluation of the magnetic field components and by analyzing the results of a test-particle simulation. In Sec.~\ref{wav}, a systematic investigation in terms of a Lyapunov ansatz is undertaken in order to determine the number of wave modes that is required to obtain a stochastic particle motion in slab turbulence. In Sec.~\ref{summ}, the results are summarized and future applications are discussed.

\section{Generating Turbulence}\label{rot}

In this section, a general method for the generation of artificial magnetic turbulence will be presented, which is based on the work of GJ99.

\subsection{General Method}

The general prescription for the numerical generation of magnetic turbulence is that of a three-dimensional, stochastic Fourier transform
\be\label{eq:dB}
\f{\hat e}_{\delta B}=\eps\sum_{n=1}^{N_m}\f\xi_n A(k_n)\cos\!\left[k_nz'+\zeta_n-\omega(k_n)t\right],
\ee
where $\eps$ is a correction factor (see Sec.~\ref{rot:dir}). Furthermore,  $\zeta_n$ is the random phase angle, and $\omega(\f k)$ is the dispersion relation if plasma-wave propagation effects are to be included (for magnetostatic turbulence, $\omega$ can be set to zero). According to GJ99, the polarization vector $\f\xi_n$ is given by
\be\label{eq:xi-GJ99}
\f\xi^{\text{GJ99}}=
\begin{pmatrix}
-i\sin\alpha\sin\phi+\cos\alpha\cos\theta\cos\phi\\
i\sin\alpha\cos\phi+\cos\alpha\cos\theta\sin\phi\\
-\sin\theta\cos\alpha
\end{pmatrix}
\ee
with additionally $\exp[i\dots]$ instead of $\cos[\dots]$ in Eq.~\eqref{eq:dB}. For simplicity we omitted the index $ n $ here but bear in mind that the angles $ \alpha, \theta, $ and $ \phi $ are different for each wavemode. The sum in Eq.~\eqref{eq:dB} extends over $N_m$ logarithmically spaced wavenumbers so that $\De k_n/k_n$ is kept constant. Throughout, normalized wavenumbers are used as $k=\ell_0\tilde k$ with $\tilde k$ the true wavenumber and $\ell_0\approx0.03$\,a.\,u. the bend-over scale.

The amplitude function, $A(k_n)$, is given through
\be\label{eq:A}
A^2(k_n)=G(k_n)\De k_n\left(\sum_{\nu=1}^{N_m}G(k_\nu)\,\De k_\nu\right)^{-1},
\ee
where $G(k)$ is the turbulence power spectrum. A useful form for $G(k)$ that allows for variable energy and inertial range spectral indices is\cite{sha09:flr}
\be\label{eq:G}
G(k)=\frac{2\left(\delta B\right)^2\ell_0\Ga\bigl((s+q)/2\bigr)}{\Ga\bigl((s-1)/2\bigr)\Ga\bigl((q+1)/2\bigr)}\,\frac{k^q}{\left(1+k^2\right)^{(s+q)/2}},
\ee
where usually $s=5/3$ is chosen to reflect the Kolmogorov inertial range.\cite{kol91:tur,bru05:sol}
Note that, due to the normalization factor in Eq.~\eqref{eq:A}, it is sufficient to use only the last factor---i.\,e., the two power-laws---of Eq.~\eqref{eq:G} when using the spectrum for the purpose of numerical turbulence generation.

Furthermore, the use of a cosine instead of combining a complex exponential function with a complex polarization vector has the additional advantage that the required computation time can be significantly decreased (by almost a factor $2$). Such is owed to the fact that, in calculating the particle trajectories, typically more than 90\% of the computation time is spent in the evaluation of Eq.~\eqref{eq:dB}. Therefore, any optimization in the implementation immediately pays off.

The connection with the magnetic turbulence strength is obtained from the condition
\be\label{eq:dB2}
\left(\delta B\right)^2=\int\df^3k\;\text{Tr}\:\mathsf P_{\ell m}(\f k),
\ee
where the trace of the magnetic correlation tensor, $\mathsf P_{\ell m}(\f k)$, is integrated. The well-known isotropic form\cite{bat82:tur,rs:rays} reads
\be\label{eq:Plm}
\mathsf P_{\ell m}(\f k)=\frac{G(\f k)}{8\pi k^2}\left(\delta_{\ell m}-\frac{k_\ell k_m}{k^2}+i\sigma(\f k)\eps_{\ell mn}\,\frac{k_n}{k}\right),
\ee
where $\sigma$ is the magnetic helicity. Inserting Eq.~\eqref{eq:Plm} in Eq.~\eqref{eq:dB2} yields
\be\label{eq:dBGk}
\left(\delta B\right)^2=\int\df^3k\;\frac{G(k)}{4\pi k^2}=\int_0^\infty G(k)\,\df k
\ee
if $G$ depends solely on the magnitude of the wave vector. Such underscores the validity of Eqs.~\eqref{eq:A} and \eqref{eq:G} without the need for a modification of the spectral index depending on the number of dimensions.\cite{gia99:sim,rev08:tra}

\subsection{A New Transformation Matrix}\label{rot:dir}

In previous implementations,\cite{tau10:pad} the rotation matrix as given through Eq.~\eqref{eq:xi-GJ99} [see Eq.~(5) of GJ99] had been used. Here, in contrast, the generation of random directions proceeds as follows. First, with $\f k=k\f\kappa$, a random wave (unit) vector is chosen as
\be
\f\kappa=
\begin{pmatrix}
\sqrt{1-\eta^2}\cos\phi\\
\sqrt{1-\eta^2}\sin\phi\\
\eta
\end{pmatrix}
\ee
with $\phi\in[0,2\pi]$ and $\cos\theta=\eta\in[-1,1]$ the cosine of the polar angle $\theta$. The parameter $\eta$ is chosen so that the density $\df\Om=\sin\theta\,\df\theta\,\df\phi=\df\eta\,\df\phi$ is constant over the entire sphere, which requirement would not be met if $\theta$ were to be uniformly distributed in the interval $[0,\pi]$.

\begin{figure}[tb]
\centering
\includegraphics[width=0.9\linewidth]{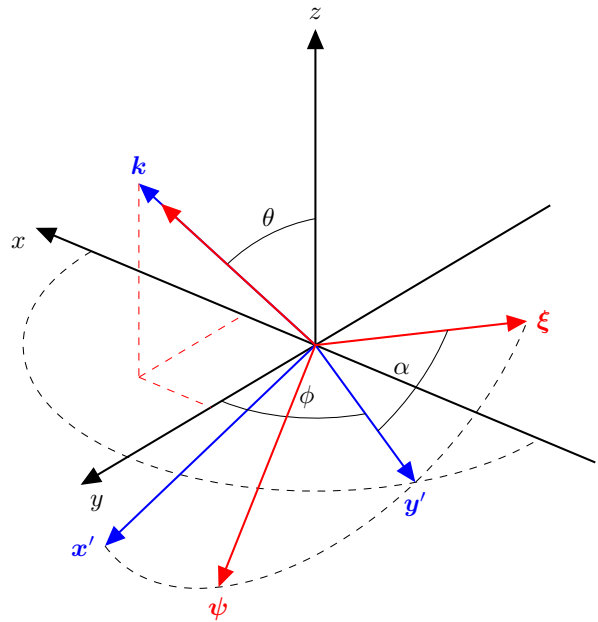}
\caption{(Color online) Illustration of the angles used for the construction of the rotation matrix. Values are used as $\phi=60^\circ$, $\theta=40^\circ$, and $\alpha=40^\circ$.}
\label{ab:3dcoord}
\end{figure}

The next step is to choose two vectors that are always perpendicular to each other and to $\f\kappa$. A straightforward choice is to restrict one vector, $\f y'$, to the $x$-$y$ plane (see Fig.~\ref{ab:3dcoord}). To complete the (right-handed) orthogonal system, set $\f x'=\f y'\times\f\kappa$ so that
\begin{align*}
\f x'&=\left(\cos\theta\cos\phi,\cos\theta\sin\phi,-\sin\theta\right)\\
\f y'&=\left(-\sin\phi,\cos\phi,0\right).
\end{align*}

Obviously, the vectors $\f x'$ and $\f y'$ are not random as $\f y'$ was restricted to the $x$-$y$ plane. Therefore, an additional random angle $\alpha\in[0,2\pi]$ is used to obtain two new vectors from a rotation in the $\f x'$-$\f y'$ plane:
\bs
\begin{align}
\f\psi&=
\begin{pmatrix}
\sin\phi\sin\alpha+\eta\cos\phi\cos\alpha\\
-\cos\phi\sin\alpha+\eta\sin\phi\cos\alpha\\
-\sqrt{1-\eta^2}\,\cos\alpha
\end{pmatrix} \label{eq:psi}\\
\f\xi&=
\begin{pmatrix}
-\sin\phi\cos\alpha+\eta\cos\phi\sin\alpha\\
\cos\phi\cos\alpha+\eta\sin\phi\sin\alpha\\
-\sqrt{1-\eta^2}\,\sin\alpha
\end{pmatrix}.
\label{eq:xi-new}
\end{align}
\es
This rotation ensures that all transformed unit vectors $ \f\psi,\f\xi,$ and $\f\kappa $ (in particular: $\ex\to\f\psi$, $\ey\to\f\xi$, and $\ez\to\f\kappa$) are perpendicular to each other and have random directions. Note that the results agree with the \emph{``y-convention''} of the Eulerian rotation matrix $\mathsf M(\f\psi,\f\xi,\f\kappa)$; the random parameters are (the Eulerian angles) $\phi$, $\eta$, and $\alpha$. Note also that the choice of Eq.~(\ref{eq:xi-new}) as the designated polarization vector $\f\xi$ is completely arbitrary. For computational/numerical purposes $ \f \psi $ and $ \f \xi $ are equivalent and, thus, interchangeable. In fact, when compared to the polarization vector of GJ99, we find that $ \f \xi^{\text{GJ99}} $ (\ref{eq:xi-GJ99}) and $ \f \psi $ (\ref{eq:psi}) are almost identical. The only difference is the imaginary number in the $x$- and $y$-components of $\f\xi^{\text{GJ99}}$ (\ref{eq:xi-GJ99}).

\begin{figure*}[tb]
\centering
\includegraphics[bb=0 316 593 525,clip,width=\linewidth]{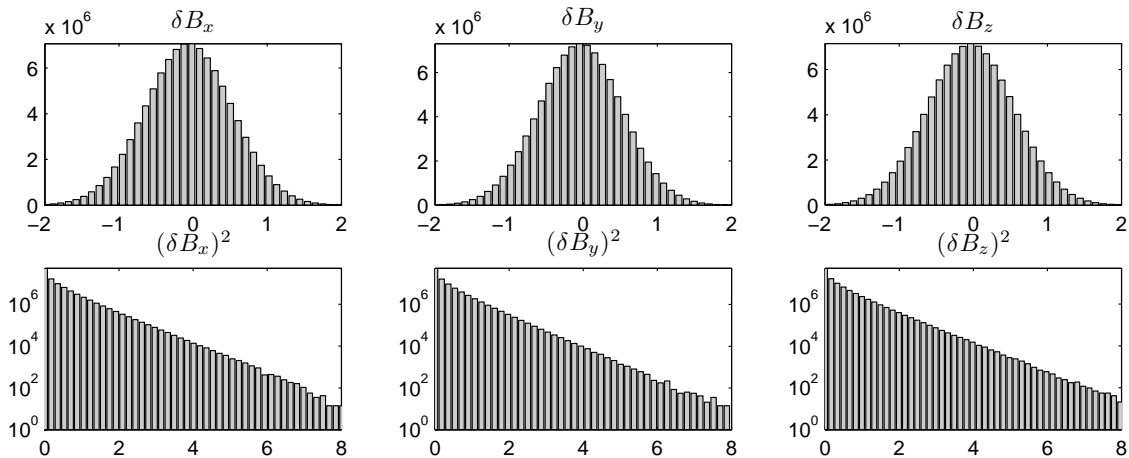}
\caption{Sampling of $10^8$ magnetic field values at random positions. 
Shown are the histograms of the $\delta B_x$, $\delta B_y$, and $\delta B_z$ values (upper panels) together with the histogram 
for the $(\delta B_x)^2$, $(\delta B_y)^2$, and $(\delta B_z)^2$ values (lower panels).}
\label{ab:magnN}
\end{figure*}

This improved transformation matrix is the first main result in this paper. The advancement of this matrix is that for the test case of isotropic turbulence all three basic requirements (see Introduction) are fulfilled simultaneously:

\setcounter{paragraph}{0}
\paragraph{Isotropy Requirement:}
The turbulence is, in fact, isotropic. Thus, all values of the polarization vector $ \f \xi $ have the same probability, so that the expectation value (mean) for each vector component vanishes,
\be
\overline{\xi_i}\equiv\frac{1}{8\pi^2}\int_0^{2\pi}\df\alpha\int_{-1}^1\df\eta\int_0^{2\pi}\df\phi\;\xi_i=0,
\ee
where $i\in\{x,y,z\}$.

\paragraph{Divergence Requirement:} The divergence of the turbulent magnetic fields is zero, $ \nabla \cdot \f B = 0 $. Thus, the turbulence is divergence free, which corresponds to $ \f k \cdot \f \xi = 0 $. This condition is always fulfilled, since $ \f k $ and $ \f \xi $ were constructed that way.

\paragraph{Magnitude Requirement:} All three turbulent magnetic field components have on average the same magnitude,
\be\label{eq:xi_ave}
\overline{\xi_i^2}\equiv\frac{1}{8\pi^2}\int_0^{2\pi}\df\alpha\int_{-1}^1\df\eta\int_0^{2\pi}\df\phi\;\xi_i^2=\frac{1}{3},
\ee
which, according to Eq.~\eqref{eq:dB}, leads to $\overline{(\delta B_i)^2}=1/3$.

At this point, we want to emphasize that all requirements are fulfilled simultaneously. This supersedes the need for different correction factors, which had to be introduced\cite{tau10:pad} for the classic GJ99 rotation matrix, where either $\overline{\xi_x^2}=\overline{\xi_y^2}\neq\overline{\xi_z^2}$ or, if that is corrected, the turbulent field is not divergence-free or, if that is corrected, the surface density for the wave vector directions is not uniform.\cite{tau12:sim}

The turbulent magnetic field vector as a function of space and time is then determined from Eq.~\eqref{eq:dB}. There, in addition to Eq.~\eqref{eq:xi_ave}, the square average of the cosine contributes a factor $1/\sqrt2$, which is remedied via the correction factor $\eps=\sqrt2$.

\subsection{Numerical Test: Results for Isotropic Turbulence}

As a quick test, we calculate numerically the mean squared values of each component of the turbulent magnetic fields and the scattering mean-free paths \emph{without} any guide magnetic field.

First, numerical sampling of $10^8$ turbulent magnetic field vectors with random spatial positions yields
\begin{align*}
\overline{(\delta B_x)^2}&=0.342\pm0.481\\
\overline{(\delta B_y)^2}&=0.314\pm0.443\\
\overline{(\delta B_z)^2}&=0.345\pm0.485.
\end{align*}
The total turbulent magnetic field strength is
\begin{equation*}
\overline{(\delta B)^2}=1.00036\pm7.67\times10^{-5}
\end{equation*}
in agreement with Eqs.~\eqref{eq:dB}, \eqref{eq:dBGk}, and \eqref{eq:xi_ave}. The distribution of $\delta B_x$, $\delta B_y$, and $\delta B_z$ together with their squares are shown in Fig.~\ref{ab:magnN}. It is thereby illustrated that all magnetic field components equally have a mean of zero (upper panels) and approximately equal variances or mean square values (lower panels), which corresponds to the first and second moments, respectively. Due to the rotational symmetry inherent in all turbulence generators discussed here, it is particularly useful to consider the behavior of the $z$ component as opposed to the $x$ and $y$ components.

\begin{figure}[tb]
\centering
\includegraphics[bb=90 260 480 565,clip,width=\linewidth]{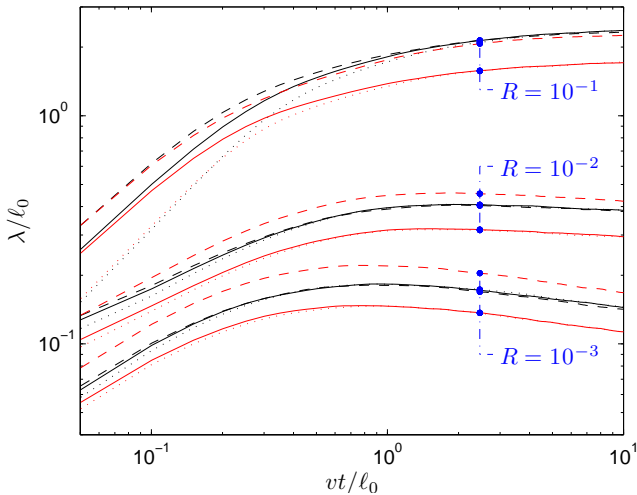}
\caption{(Color online) The three running mean-free paths (solid lines: $\lambda_x$; dotted lines: $\lambda_y$; dashed lines: $\lambda_z$) for the rotation matrix described in Sec.~\ref{rot} (black lines) as opposed to the previously used version\cite{gia99:sim,tau10:pad} (red lines). The parameter $R=\gamma v/(\ell_0\Om)$ denotes the normalized rigidity, which is proportional to the particles' velocity.}
\label{ab:isomfp}
\end{figure}

Second, Fig.~\ref{ab:isomfp} shows the results for the ``running'' mean-free paths,
\be
\lambda_i(t)=\frac{3}{2vt}\m{\left(\De r_i\right)^2},\qquad i\in\{x,y,z\},
\ee
where the angular brackets $\langle\cdots\rangle$ denote the averaging over -- in this case -- an ensemble of $10^5$ particles.

As expected, in the long time limit $\lambda_x=\lambda_y=\lambda_z$ (black lines). Furthermore, three different particle energies are assumed via the normalized rigidity, $R=\gamma v/(\ell_0\Om)$. While for small times, there are some deviations between the three components, there is clear convergence towards a common value for large times.

In contrast, the GJ99 rotation matrix\cite{gia99:sim,tau10:pad} exhibits a clear difference between the mean-free paths. For all energies considered we find that $ \lambda_z>\lambda_x=\lambda_y$. Somewhat interestingly, for almost all particle energies (except high energies) the new, corrected, mean free path lies between $\lambda_z^{\text{GJ99}}$ and $\lambda_{x,y}^{\text{GJ99}}$. However, for high-energetic particles we find that $\lambda_z$ was underestimated, while $\lambda_x$ and $\lambda_y$ are almost in agreement with the new, common value.

\subsection{Special Geometries}

The above considerations are valid for the cases of isotropic and continuous anisotropic turbulence geometries, where the latter is for example given by the Maltese cross geometry.\cite{wei10:mal,rau12:mal} However, several additional standard cases have to be considered.

\setcounter{paragraph}{0}
\paragraph{Slab Geometry:} Slab turbulence is characterized by $\delta\f B(\f r,t)=\delta\f B(z,t)$, so that $\f k=k\ez$ and $\delta\f B\perp\ez$. 
Hence, $\eta$ has to be set to unity but otherwise the above considerations remain unchanged.
Without loss of generality we set $ \alpha=0 $, so that the vectors can be rewritten as
\bs
\begin{align}
\f \psi &= \left( \cos \phi,\sin \phi,0 \right)  	\\
\f\xi &= \left( -\sin \phi,\cos \phi,0 \right)  	\\
\f\kappa&= \left( 0,0,1 \right).
\end{align}
\es

\paragraph{2D Geometry:} The opposite case of two-dimensional turbulence is characterized via $\delta\f B(\f r,t)=\delta\f B(x,y,t)$ so that $\f k\perp\ez$. In addition,\citep{gra96:sca,bie96:two} one usually requires $\delta\f B\perp\ez$. Therefore, without loss of generality one sets $\eta=0$ and $\alpha=0$ so that
\bs
\begin{align}
\f \psi &= \left( 0,0,-1 \right)  	\\
\f\xi &=\left(-\sin\phi,\cos\phi,0\right) 		\\
\f\kappa&=\left(\cos\phi,\sin\phi,0\right).
\end{align}
\es
If one were to ignore the condition that $\delta\f B$ be perpendicular to the background field, $\alpha$ may also be chosen randomly.

\paragraph{Composite Geometry:} Slab and 2D geometries can be combined to obtain a quasi-three-dimensional magnetic field as
\be
\delta\f B_{\text{comp}}(\f r,t)=\delta\f B_{\text{slab}}(z,t)+\delta\f B_{\text{2D}}(x,y,t),
\ee
which is named composite turbulence. Here, care has to be taken that the amplitude functions are modified so that $A_{\text{slab}}(k)\to A_{\text{slab}}(k)r$ and $A_{\text{2D}}(k)\to A_{\text{2D}}(k)(1-r)$, where $r$ describes the relative weight of the slab contribution according to $(\delta B)^2=r(\delta B_{\text{slab}})^2+(1-r)(\delta B_{\text{2D}})^2$. Typical values\cite{bie94:pal,bie96:two} vary between $r=15\%$ and $r=20\%$.

\subsection{Electric Fields}

\begin{figure*}[tb]
\centering
\includegraphics[width=0.32\linewidth]{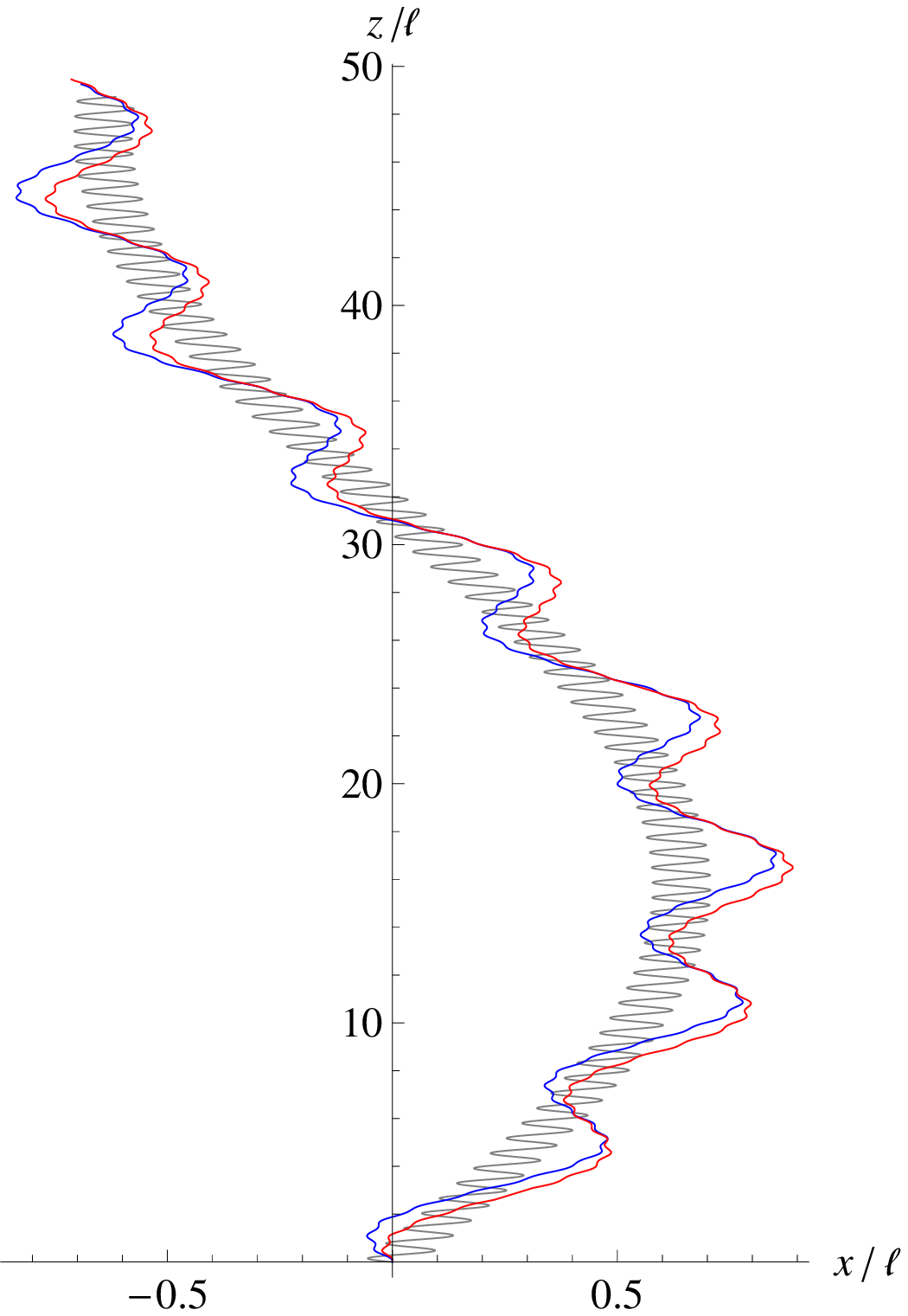}
\includegraphics[width=0.32\linewidth]{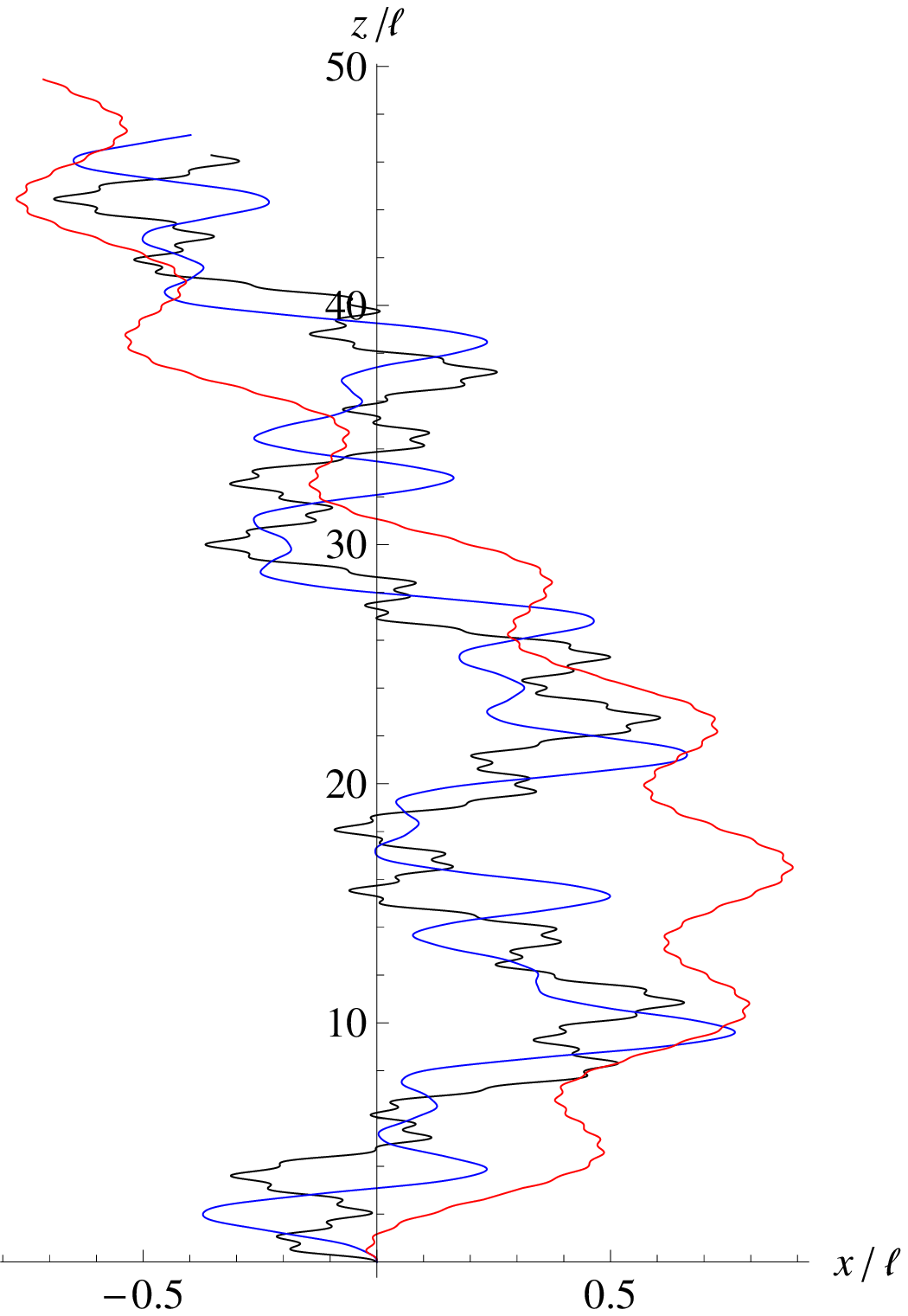}
\includegraphics[width=0.32\linewidth]{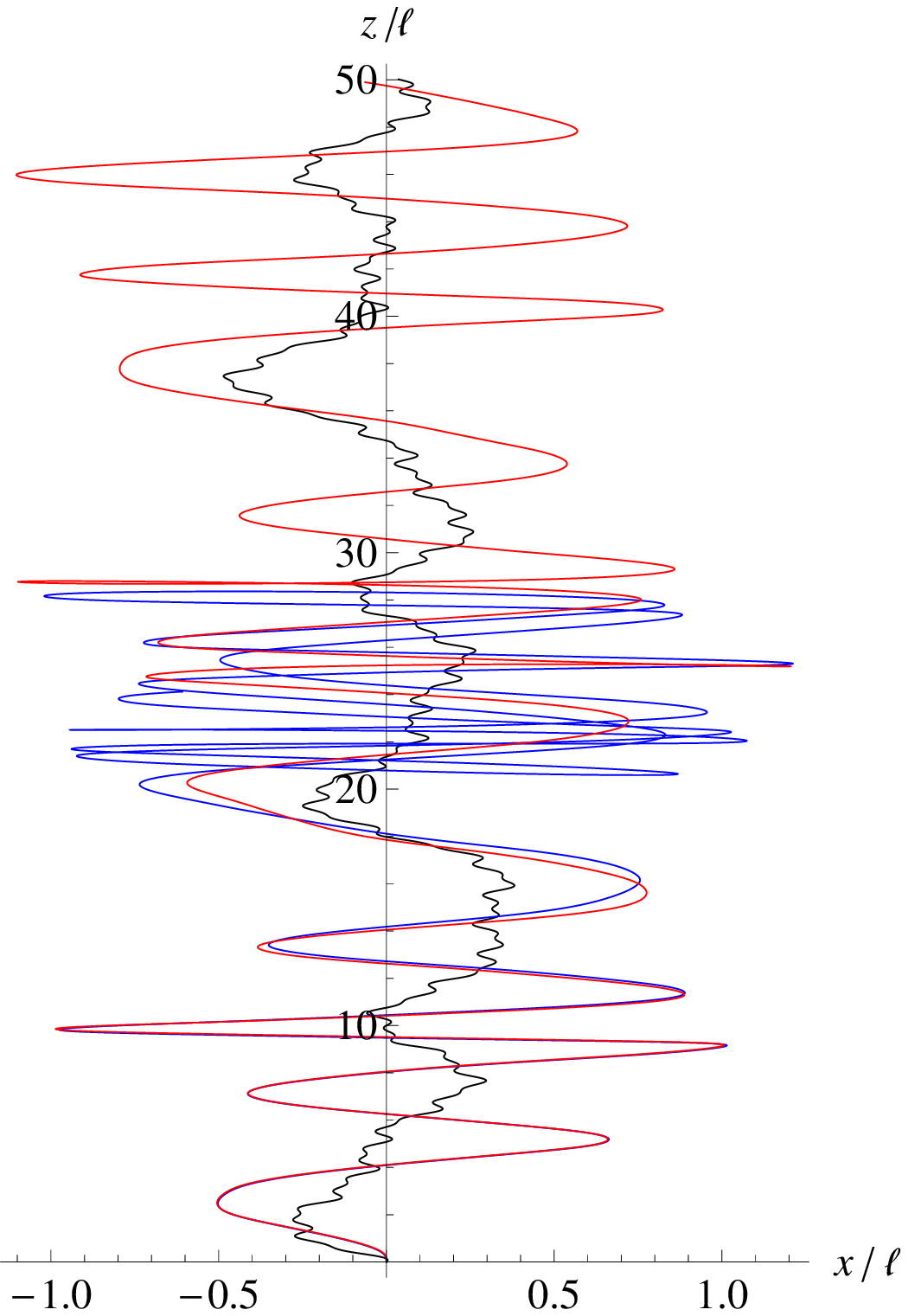}
\caption{(Color online) Trajectories of two particles starting at $z=0$ (blue lines) and $z=0.1\,\ell_0$ (red lines) together with a magnetic field line (black line). From left to right, the three panels show the cases of two, four, and eight wave modes are shown, corresponding to the value for the parameter $N_m$ in Eq.~\eqref{eq:dB}.}
\label{ab:traj}
\end{figure*}

If plasma waves are to be included via the dispersion relation $\omega(\f k)$ in Eq.~\eqref{eq:dB}, turbulent electric fields are induced via Faraday's law, which (i) modify the diffusion coefficients and (ii) lead to stochastic acceleration\citep{tau10:wav,tau12:sto} via momentum diffusion.\citep{mic96:alf,ost97:sof,rs:rays,sta08:mom} In general, the solution for $\f E$ of the Faraday equation $\f B=(c/\omega)\f k\times\f E$ is ambiguous. However, for basic magnetohydrodynamic (MHD) waves such as (fast) magnetosonic waves and Alfv\'en waves, the properties of the electric field vector and its amplitude are known.\citep{rs:rays} In turn, the two MHD wave types will be discussed.

As a side note, there are subtle issues about the use of random phase factors in Eq.~\eqref{eq:dB} for the determination of particle acceleration. While the assumption of randomized phases is frequently used in test-particle simulations as well as for diffusive-shock acceleration, nonlinear wave-wave interaction processes tend to generate coherence among Alfv\'en waves.\cite{arz06:coh,mat09:acc} Therefore, the best way of generating the proper electromagnetic field structures would be to solve the magnetohydrodynamic equations.\cite{tea08:tra}

\subsubsection{Fast Magnetosonic Waves}

The relevant properties of the electric field unit vector, $\f{\hat e}_{\delta E}$, are: (i) $\f{\hat e}_{\delta E}\perp\f{\hat e}_{\delta B}$; (ii) $\f{\hat e}_{\delta E}\perp\f k$; and (iii) $\f{\hat e}_{\delta E}$ is independent of \bo. Combining the second condition with Faraday's law, the electric field strength is given by
\be
\abs E=\frac{\vA}{c}\abs B,
\ee
where the dispersion relation $\omega=\pm\vA k$ has been used with $\vA=B_0/\sqrt{4\pi\rho}$ the Alfv\'en speed.

Because $\f k\times\f\psi=\f\xi$, the orientation of the electric field is simply that of the vector $\f\psi$ as given in Eq.~\eqref{eq:psi}.

\subsubsection{Alfv\'en Waves}

Here, the geometric properties are slightly modified, insofar as: (i) $\f{\hat e}_{\delta E}\perp\f{\hat e}_{\delta B}$; (ii) $\f{\hat e}_{\delta E}\cdot\f k=k\sin\theta$ with $\theta=\angle\left(\f k,\bo\right)$ as in Sec.~\ref{rot:dir}; and (iii) $\f{\hat e}_{\delta E}\perp\bo$ so that the $z$ component is automatically zero.


Together with Faraday's law, the second condition implies that, for $\theta\to\pi/2$, the magnetic field decreases and, at the same time, the frequency decreases because of the dispersion relation $\omega=\pm\vA k\cos\theta$. Such could be taken into account by setting
\bs
\begin{align}
\f\xi&=\left(-\cos\theta\sin\phi,\cos\theta\cos\phi,0\right)\\
\f\psi&=\left(\cos\phi,\sin\phi,0\right).
\end{align}
\es
In that case, the correction factor has to be set to $\eps=\sqrt6$ because $\overline{(\delta B_{x,y})^2}=1/6$. However, it is undesirable to have a turbulent magnetic field strength that varies with the direction and is therefore reduced on average. Therefore, it is useful to restrict the wave vector to the $z$ direction (i.\,e., set $\theta=0$) so that slab geometry is retained.

\section{Number of Wavemodes}\label{wav}

\begin{figure}[tb]
\centering
\includegraphics[width=0.95\linewidth]{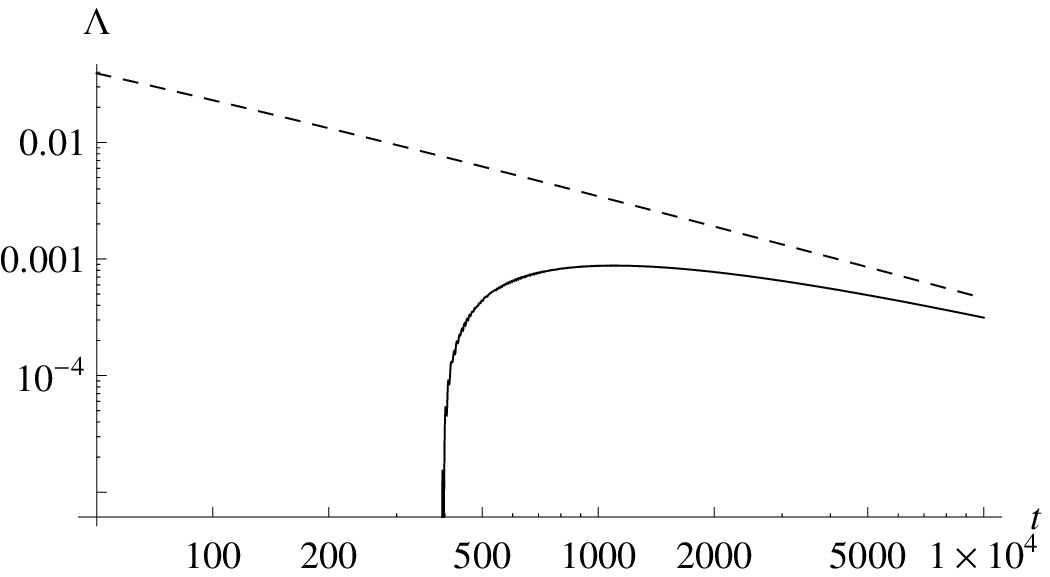}\\
\includegraphics[width=0.95\linewidth]{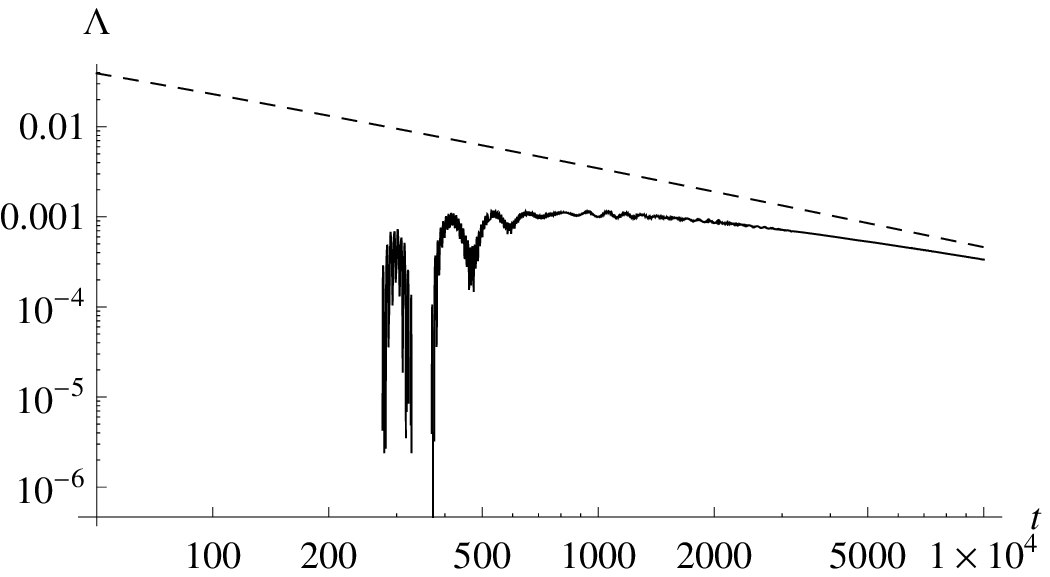}\\
\includegraphics[width=0.95\linewidth]{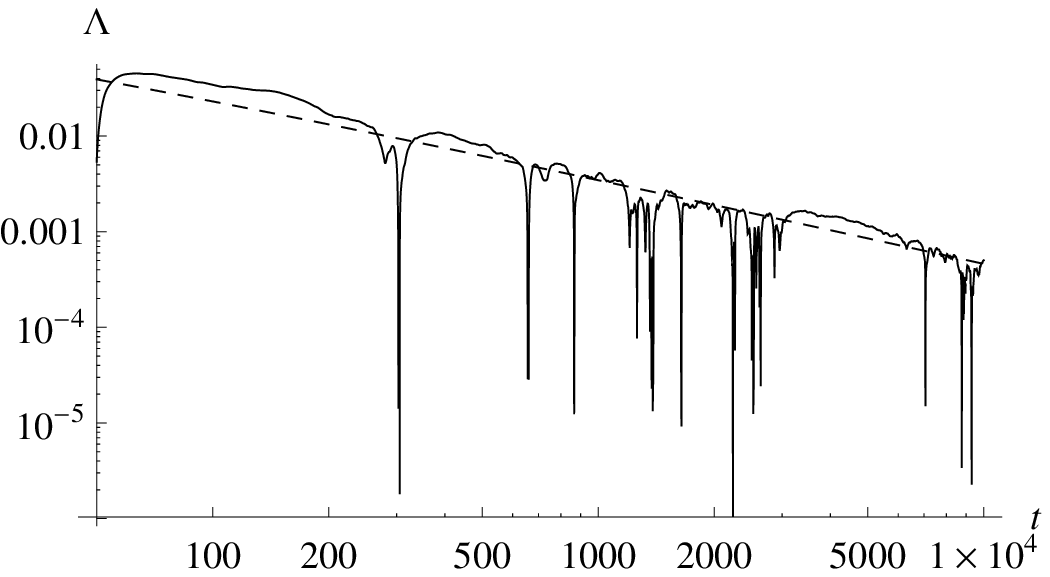}
\caption{The quasi-Lyapunov exponent from Eq.~\eqref{eq:lya} as obtained from the time evolution of the spatial deviation between two initially neighboring particles. From top to bottom, the three panels show the cases of two, four, and eight wave modes are shown. Note that, in the upper and middle panels, $\La$ is negative for early times $t\lesssim300$.}
\label{ab:lya}
\end{figure}

A second, related problem is the following: How many wave modes (i.\,e., what value has to be chosen for $N_m$) are needed so that Eq.~\eqref{eq:dB} describes a \emph{turbulent} magnetic field in a strict physical sense? In what follows, first a criterion is found which is then put to the test by running a full simulation.

\subsection{Simplified Turbulence}\label{wav:simp}

To shed some light on this matter, a simplified equation of motion, $\f{\dot v}=(q/c)\f v\times\f B$ is solved numerically for a magnetic field of the form
\be
\f B_{\{x,y\}}(\f r,t)=\sum_{n=1}^{N_m}A(k)
\begin{Bmatrix}
\cos\\
\sin
\end{Bmatrix}
(\alpha_n)\cos(kz)
\ee
with the amplitude function from Eq.~\eqref{eq:A} and with $\alpha_n=[2(n-1)/N_m+1/4]\pi$. For a single wave mode, the particle motion remains regular. That behavior is already modified if $N_m=2$. In Fig.~\ref{ab:traj} the trajectories of two neighboring particles for two, four, and eight wave modes are illustrated together with a magnetic field line.

The question whether or not particles are ``randomly'' scattered is investigated by observing the time evolution of the distance $\De z$ between two initially neighboring particles (with a distance of, say, $\De z=0.1\,\ell_0$). While, $\De z$ remains constant for one wave mode, an approximately linear increase is found for the cases of two and four wave modes as exhibited by comparison of the two trajectories in Fig.~\ref{ab:traj}. That behavior is entirely changed for eight wave modes, when a stochastic variation is found between regimes with positive and negative $\De z$.

A more rigorous mathematical criterion is found in chaotic systems, where initially neighboring instances deviate exponentially according to
\bs
\be
\De z(t)\approx\De z(0)\exp\left(\La t\right),
\ee
where
\be\label{eq:lya}
\La=\lim_{t\to\infty}\lim_{\De z(0)\to0}\,\frac{1}{t}\,\ln\abs{\frac{\De z(t)}{\De z(0)}}
\ee
\es
is a quasi-Lyapunov exponent. However, here the total particle velocity is constant so that an exponential deviation would be possible only for small times. Furthermore, a truly random behavior leads to \emph{diffusion}, in which case one expects $\De z\propto\sqrt t$. Therefore, a quasi-Lyapunov exponent should exhibit a behavior $\La\propto\ln(t)/t$ if $N_m$ is sufficiently large so that particles diffuse. As shown in Fig.~\ref{ab:lya} for the cases of two, four, and eight wave modes, that requirement is almost (but not quite) fulfilled for two and four wave modes, while, for eight wave modes, an excellent agreement is seen even for small times. The fact that $\La$ becomes very small at intermittent time intervals underscores that, from time to time, particles can move toward each other, as is expected from a random-walk motion.

\begin{figure}[t]
\centering
\includegraphics[bb=98 190 475 635,clip,width=\linewidth]{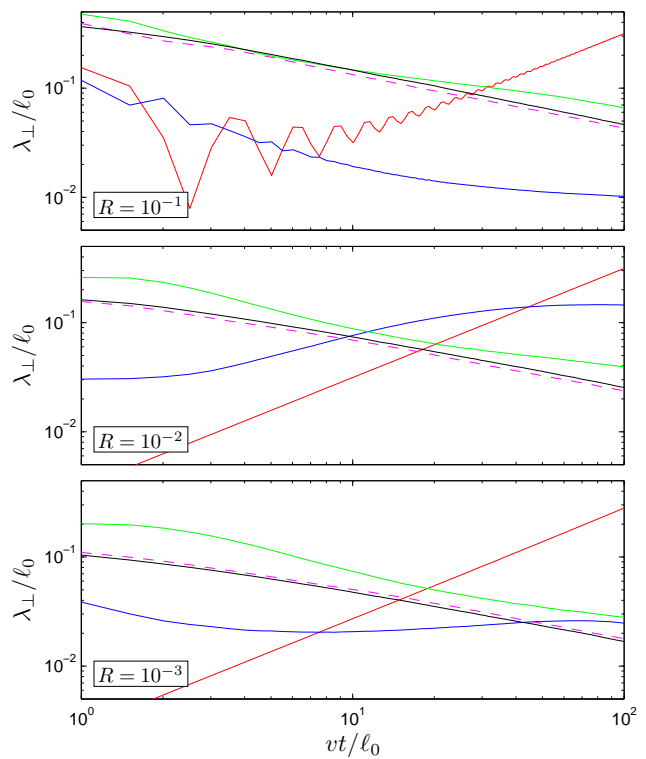}
\caption{(Color online) Perpendicular mean-free path, $\lambda\se/\ell_0$, as a function of the normalized simulation time $vt/\ell_0$. Shown are the cases of two (red), four (blue), eight (green), 16 (magenta dashed), and 512 (black) wave modes. Note that additional runs with 32 and 128 wave modes are almost indistinguishable from the case of 512 wave modes.}
\label{ab:wavmod_ls}
\end{figure}

\begin{figure}[t]
\centering
\includegraphics[bb=98 190 475 635,clip,width=\linewidth]{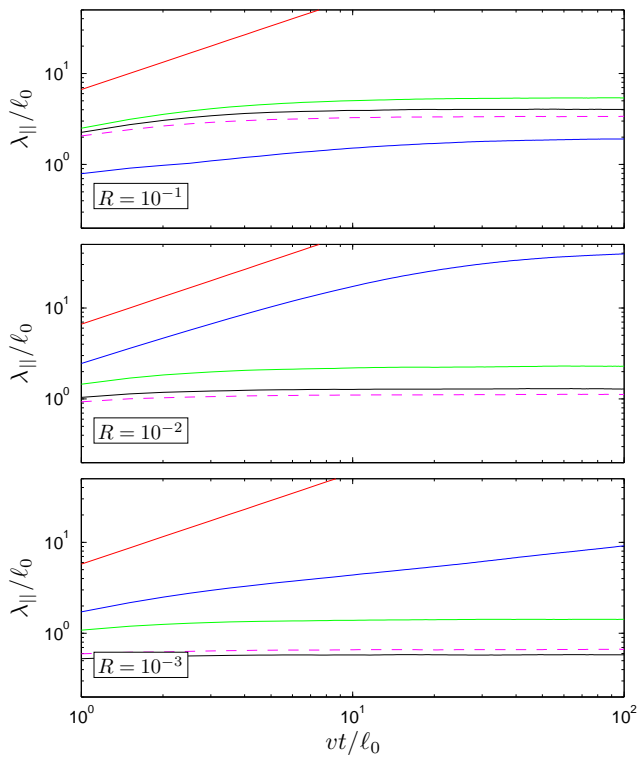}
\caption{(Color online) Same as Fig.~\ref{ab:wavmod_ls}, only that here the parallel mean-free path, $\lambda\pa/\ell_0$, is shown.}
\label{ab:wavmod_lz}
\end{figure}

\subsection{Full Turbulence Simulation}

To verify the above findings, a series of simulations has been run using the \textsc{Padian} code\cite{tau10:pad} for slab turbulence of the average strength $\delta B=B_0$. The resulting perpendicular and parallel mean-free paths are shown in Figs.~\ref{ab:wavmod_ls} and \ref{ab:wavmod_lz}, respectively, which have been obtained from averaging over $10^5$ particles in $250$ turbulence realizations.

The number of wave modes, $N_m$, as used for the evaluation of Eq.~\eqref{eq:dB} is varied between $N_m=2$ and $N_m=512$. As expected from the previous considerations, diffusive behavior is seen for eight wave modes. Both the parallel and the perpendicular running diffusion coefficients exhibit the same behavior that is found for a considerably higher number of wave modes. However, the absolute value of the mean-free paths still deviate significantly from the true values.

In agreement with the results found in Sec.~\ref{wav:simp}, the overall conclusion is that eight wave modes are sufficient for a diffusive particle behavior. The reason is that such allows for particle trapping (as exhibited in the right-hand panel of Fig.~\ref{ab:traj} at $z=20\,\ell_0$), which however depends on the precise initial conditions so that some particles become trapped with a subsequent reversal of their parallel motion. Of course, for other than slab turbulence, one expects that more wave modes are necessary so that the three-dimensional structures are sufficiently resolved. Additionally, if different wavenumber ranges contribute to the spectrum (such as energy, inertial, and dissipation ranges\cite{bru05:sol}), each range has to be sampled with its own set of wavenumbers.

\section{Discussion and Conclusion}\label{summ}

In this article, the generation of numerical electromagnetic turbulence has been revisited. Such turbulent fields are frequently used in test-particle simulations to determine the scattering parameters of energetic particles such as cosmic rays. Several methods use the approach of superposing a number of randomly propagating Fourier modes, which is theoretically well-based.\cite{bat82:tur} Whereas previous approaches sometimes lacked the required randomness of the Fourier modes as well as the properties of the resulting turbulent fields, here a new derivation has been presented. Therefore, it has to be pointed out that the new derivation presented here solves all three problems explained in the introduction:\cite{tau12:sim} (i) all three turbulent magnetic field components have on average the sample magnitude, i.\,e., $\overline{(\delta B_i)^2}=1/3$; (ii) if all parameters are chosen random, the turbulence is precisely isotropic so that a vanishing mean of all turbulent magnetic field components is found, i.\,e., $\overline{\delta B_i}=0$; and (iii) the turbulence is divergence-free, which, in Fourier space, corresponds to $\f k\cdot\f\xi=0$ and is fulfilled by construction.

In addition, the absolute strength of the turbulent magnetic field plays an important role not only for the magnitude of the (both running and final, diffusive) mean free path but also for the entire particle behavior as the underlying scaling laws are controlled by the Kubo number,\cite{isi92:per,zim09:per} which is defined as $K=(\delta B/B_0)/(\ell_\parallel/\ell_\perp)$ if---unlike here---the parallel and perpendicular correlations lengths differ. While previous implementations sometimes corrected the transport parameters afterwards, the correct advance is to ensure the appropriate turbulence strength prior to integrating the particle trajectories even if electric fields are included, which sometimes reduce the turbulence strength even more due to an additional sine factor.\cite{mic96:alf,tau10:wav}

Furthermore, the problem of the number of Fourier modes required for a sufficiently turbulent magnetic field has been tackled from the view point of the particle behavior. It is known that, for slab turbulence, the running perpendicular mean free path exhibits a power-law as $\lambda\se\propto t^{-1/2}$, while the parallel mean free path is expected to be diffusive.\cite{tau10:sub} Such a behavior is already seen for eight Fourier modes [i.\,e., $N_m=8$ in Eq.~\eqref{eq:dB}]. It may seem surprising at first glance that only (figuratively speaking) a handfull of wave modes is sufficient to obtain accurate results. But such is in agreement with early simulations,\cite{ost93:mom,mic96:alf,gia94:mul} which, despite the low number of Fourier modes, gave results that were nevertheless accurate enough to allow for successful comparisons with analytical theories.

In many cases one expects previous simulation results to retain their validity due to the robustness of the particle transport parameters for certain parameter regimes. In general, however, such is not guaranteed for the reasons stated above. It is reasonable to assume that each additional feature---such as anisotropic correlation lengths and/or spectral indices, a dissipation range, or time dependence---will require additional modes so that the spectrum can be sufficiently resolved. Future work should therefore attempt to find criteria as to how many wave modes are required depending on the complexity of the turbulence power spectrum.

\begin{acknowledgments}
The authors acknowledge fruitful discussions with Fathallah Alouani Bibi and Gary Zank. This work was done during RCT's stay at the University of Alabama in Huntsville, which was funded by the \emph{Deut\-sche Aka\-de\-mi\-sche Aus\-tausch\-dienst} (DAAD). The work of AD is supported by the \emph{Deutsche Forschungsgemeinschaft} (DFG) under grant DO~1505/1-1.
\end{acknowledgments}




\end{document}